\documentclass[doublecol]{epl2}
\usepackage{graphicx}
\usepackage{color}
\usepackage{amssymb}
\usepackage{amsmath} 
\usepackage[T1]{fontenc}
\usepackage{graphicx}

\usepackage[colorlinks,bookmarks=false,citecolor=blue,linkcolor=red,urlcolor=blue]{hyperref}

\title{Few-body perspective on fermionic pairing in one spatial dimension}

\author{Tomasz Sowi\'nski}
\shortauthor{T. Sowi\'nski}
\institute{Institute of Physics, Polish Academy of Sciences, Aleja Lotnikow 32/46, PL-02668 Warsaw, Poland}

\abstract{In this perspective we discuss recent theoretical and experimental concepts giving a route to a better understanding of conventional and unconventional pairing mechanisms between opposite-spin fermions arising in one-dimensional mesoscopic systems. With special attention, we focus on the problem of experimental detectability of correlations between particles. We argue that state-of-the-art experiments with few ultracold fermions may finally break an impasse and give pioneering and unquestionable verification of the existence of correlated pairs with non-zero center-of-mass momentum.
\vspace{1cm}
}

\begin{document} 

\maketitle

{\bf Introduction.} -- Typically, quantum matter exhibits its unusual properties in a very elusive way. To capture them and expose their non-classicality one needs to perform very subtle and precise measurements with highly non-trivial experimental setups. However, for over 110 years we have known the purely quantum phenomenon of {\it the superconductivity}\cite{1911Onnes} that is macroscopically displayed by a plethora of different materials cooled down to sufficiently low temperatures. On a phenomenological level, the description of superconductivity is relatively simple -- below a certain temperature the electric resistance of a material rigorously vanishes, thus an electric current can flow without any external voltage. However, the theoretical explanation of this astonishing fact is not simple at all. It took almost 50 years to find an appropriate, standing on fundamental quantum mechanical grounds, understanding of this slightly uncomplicated experimental fact \cite{2010CooperFeldmanBook} and to formulate {\it the theory of superconductivity} by Bardeen, Cooper, and Schrieffer \cite{1957BardeenPhysRev} (BCS). The BCS theory is based on a brilliant prediction of Cooper \cite{1956CooperPhysRev} that mutual attractions between opposite-spin fermions force the system to rebrand its many-body ground state and lower the ground-state energy by the collective formation of non-classically correlated pairs of opposite spin fermions. Quickly, it turned out that the pairing mechanism predicted by Cooper plays a fundamental role in many different quantum many-body systems, from heavy nuclei \cite{1958BohrPhysRev,1959MigdalNucPhys}, through ultracold gasses \cite{2008GiorginiRevModPhys}, to neutron stars\cite{1969BaymScience,1971YangNucPhysA}.\footnote{In the case of metallic crystals, it is still highly non-trivial to explain how in a repulsive system of electrons mutual attractive interactions can effectively emerge. This is another cornerstone of the BCS theory not discussed here.}

The Cooper mechanism is only the simplest way of forming correlated pairs of opposite-spin fermions and, in fact, it is supported only in systems close to perfect balance, {\it i.e.}, when Fermi surfaces of both components are identical. Then, without violation of the conservation of energy, single-particle excitations from Fermi surfaces match and may lead to bounded Cooper pairs with vanishing center-of-mass momentum. Contrarily, in the case of imbalanced systems, as predicted independently by Fulde and Ferrel \cite{1964FuldePhysRev} and Larkin and Ovchinnikov \cite{1965LarkinJETP} (FFLO), pair formation requires an additional relative shift of Fermi spheres and thus leads directly to the formation of exotic pairs with non-zero net momentum. Importantly, these predictions are to some extend robust to the origins of the imbalance, {\it i.e.}, the reasoning is similar for systems with particle imbalance, mass imbalance, or with different single-particle excitation spectra. Thus, the FFLO pairing seems to be as fundamental as the original concept of Cooper. On the experimental side, however, in contrast to standard BCS pairing, there is still a lack of direct evidence for the formation of the FFLO phases. It is believed that the most promising systems in which any tracks of the FFLO pairing can be detected are one-dimensional systems of fermionic atoms \cite{2007OrsoPRL,2010ZapataPRL,2010LiaoNature,2018KinnunenRPP,2020DobrzynieckiAdvQTech}.

In all the original argumentations described above, it is almost always assumed that the fermionic system contains a macroscopically large number of particles. Therefore, the formation of pairs, although occurring mainly close to the Fermi surfaces, is viewed as a collective bulk behavior of a system. Still open and interesting questions arise when the number of particles is essentially finite, {\it i.e.}, when observable system's properties are sensitive to varying particle number by one \cite{2017HofmannPRB,2015ChengNature}. Answering them, even partially, may significantly change our understanding of correlations in many-body systems and the way they are collectively built when the number of particles is succesively increased.

In this perspective we aim to push forward the discussion and inspire further exploration of this still little-examined issue. We discuss recent theoretical concepts which can be applied to any few-fermion system. They allow the capture of different properties of paired phases in systems containing several particles and give a consistent approach to capture their dependence on the number of particles. Although the presented concepts are very general, we want to keep in mind all experimental limitations. Therefore we focus mainly on fermionic few-body systems currently attainable in state-of-the-art experiments with ultracold gases \cite{2008CheinetPRL,2013WenzScience,2019SowinskiRPP}.

{\bf The system.} -- 
In this perspective we focus on one-dimensional, well-confined two-component fermionic mixtures containing several particles. In principle particles belonging to different components $\sigma\in\{\uparrow,\downarrow\}$ may have different masses $m_\sigma$ and may experience different external confinements $V_\sigma(x)$. For simplicity we assume that mutual interactions are present only between opposite-spin particles and they have contact character controlled by a single parameter $g$. With this limitations, after introducing fermionic field operators $\hat{\psi}_\sigma(x)$, the most general Hamiltonian of the system written in the second-quantization formalism reads
\begin{align}
\hat{H}
&= \sum_\sigma\int\!dx\,\,\hat{\psi}_\sigma^\dagger(x)\left[-\frac{\hbar^2}{2m_\sigma}\frac{d^2}{dx^2}+V_\sigma(x)\right]\hat{\psi}_\sigma(x)\nonumber \\ 
&+g\int\!dx\,\, \hat{n}_\uparrow(x)\hat{n}_\downarrow(x), \label{Hamiltonian}
\end{align}
where $\hat{n}_\sigma(x)=\hat{\psi}^\dagger_\sigma(x)\hat{\psi}_\sigma(x)$ is the density operator for $\sigma$ particles. It is clear that the Hamiltonian \eqref{Hamiltonian} commutes with the particle number operators $\hat{N}_\sigma=\int\!dx\,\,\hat{n}_\sigma(x)$. Therefore, in the following we analyze system properties in the subspaces of a fixed number of particles. To make discussion as clear as possible we restrict the discussion to the many-body ground-state $|\mathtt{g}\rangle$ of the Hamiltonian $\hat{H}$ and its energy ${\cal E}_0$. 

\begin{figure}
\includegraphics[width=0.9\linewidth]{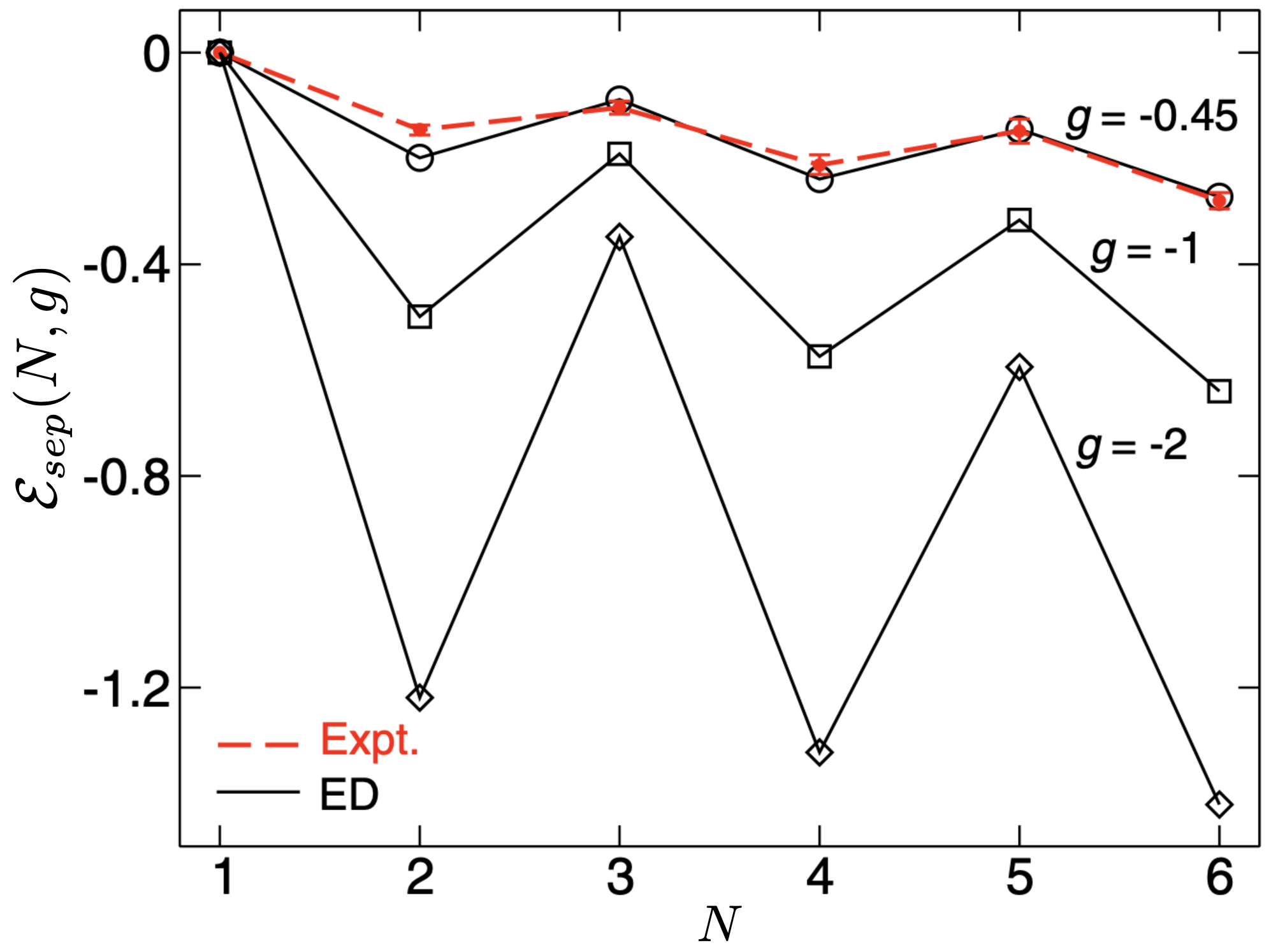}
\caption{Separation energy ${\cal E}_{sep}(N,g)$ for different number of particles and different attractive interactions. A difference between systems containing an even and odd number of fermions is evident. Collective participation of all particles is manifested by the fact that consecutive minima become deeper while maxima do not saturate at zero. Note an almost perfect agreement of theoretical predictions (black points) with experimental results reported in \cite{2013ZurnPRL} (red points). Separation energy is expressed in natural units of energy, $\hbar\Omega$. Figure reprinted with permission from \cite{2015DAmicoPRA}. Copyright (2015) by the American Physical Society.\label{Fig1}} 
\end{figure}
\begin{figure}
\includegraphics[width=\linewidth]{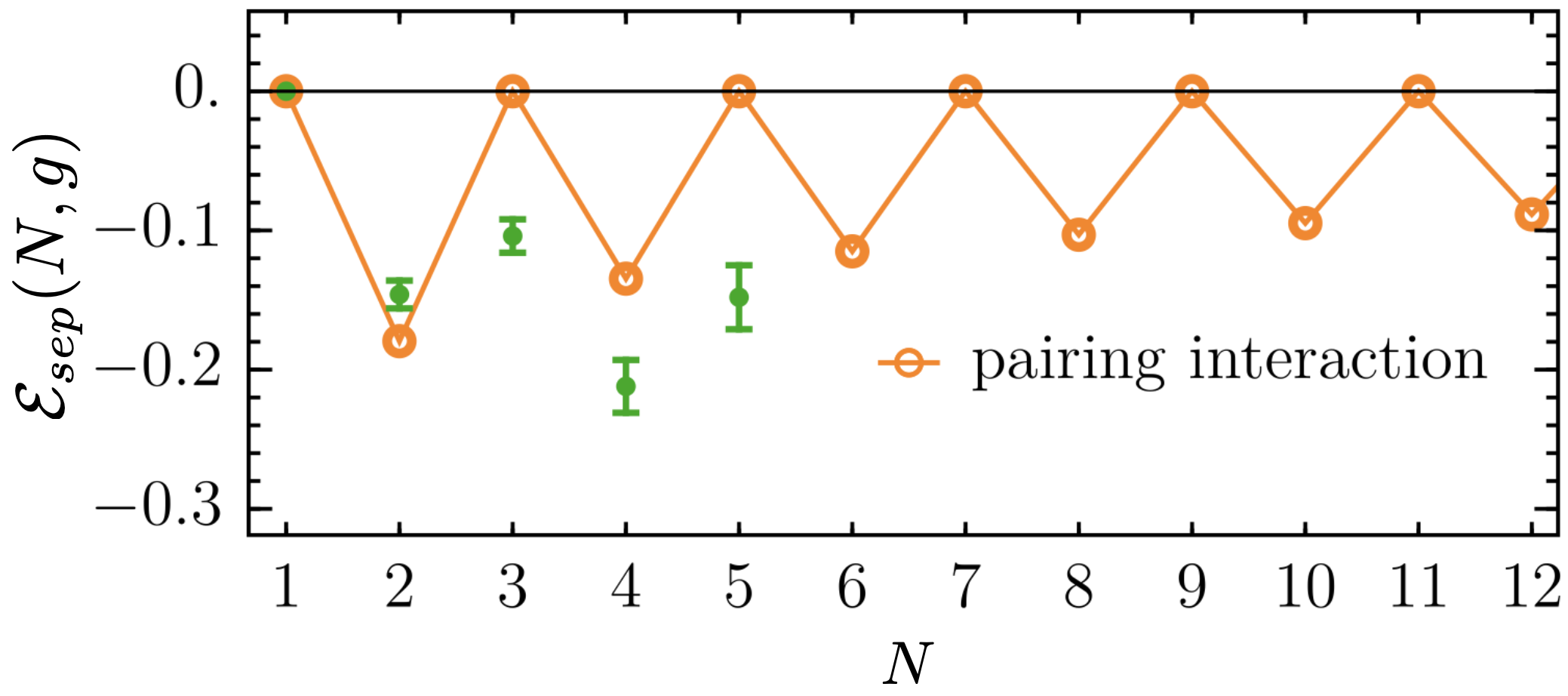}
\caption{Dependence of the separation energy ${\cal E}_{sep}(N,g)$ on the number of particles $N$ when interactions are artificially limited only to the intra-shell pairing terms. In this case, the separation energy vanishes for odd-$N$ systems and has a monotonically decreasing magnitude for successive even particle numbers. This prediction strongly contradicts experimental data (green points) indicating collective enhancement of pairing in the system. Separation energy is expressed in natural units of energy, $\hbar\Omega$. Figure reprinted with permission from \cite{2016HofmannPRA}. Copyright (2016) by the American Physical Society.\label{Fig2}} 
\end{figure}

 
{\bf Energetic considerations.} -- First signs suggesting that attractive systems described by the Hamiltonian \eqref{Hamiltonian} may manifest non-trivial collective correlations between opposite-spin fermions are encoded already in the many-body ground-state energy. To reveal these circumstances let us follow first experimental \cite{2013ZurnPRL} and theoretical \cite{2015BergerPRA,
2015DAmicoPRA,2016McKenneyJPB} considerations in this spirit and consider the simplest situation of equal mass particles ($m_\downarrow=m_\uparrow=m$), confined in a harmonic trap ($V_\sigma(x)=m\Omega^2x^2/2$), and being close to the balanced scenario. The latter means that along with increasing total number of particles $N=N_\uparrow+N_\downarrow$ one finds $\Delta N=N_\uparrow-N_\downarrow=0$ or $1$ for $N$ even or odd, respectively. It is clear that the ground-state energy ${\cal E}_0(N,g)$ is directly related to the number of particles and to the interaction strength $g$. For a fixed number of particles, it can be quite naturally interpreted as a sum of the energy of the non-interacting system ${\cal E}_0(N,0)$ and the rest which we associate with mutual interactions -- the interaction energy ${\cal E}_{int}(N,g)$. One can naively suspect that the interaction energy should be a monotonic function of the number of particles since along with increasing $N$ one finds an increasing number of opposite-spin pairs willing to interact. In fact this is not the case since the separation energy ${\cal E}_{sep}(N,g) = {\cal E}_{int}(N,g) - {\cal E}_{int}(N-1,g)$, {\it i.e.}, the difference between interaction energies of consecutive particle numbers\footnote{Note that the separation energy ${\cal E}_{sep}(N,g)$ can be equivalently expressed as a difference ${\cal E}_{sep}(N,g)=\mu(N,g)-\mu(N,0)$, where $\mu(N,g) = {\cal E}_0(N,g) - {\cal E}_0(N-1,g)$ is a finite-system version of the chemical potential quantifying energy needed to add (remove) a particle to (from) the system at given interaction strength. In this way, an additional correspondence to many-body systems can be established.}, reveals specific oscillations between even and odd particle numbers \cite{2015DAmicoPRA} (see Fig.~\ref{Fig1}). It means that even configurations (exact balance between components) have lower binding energies than neighboring odd ones. Therefore they are rather more stable. The physical mechanism behind this behavior is quite understandable -- the most important contribution to the interaction energy comes from intra-shell configurations, {\it i.e.}, from interactions between opposite-spin fermions occupying the same single-particle orbital of the external confinement. Whenever the system is imbalanced, some fermions do not have opposite-spin partners to attract and therefore to effectively decrease the system's energy. However, an explanation based on purely intra-shell interactions is not sufficient to explain a general trend, clearly visible in Fig.~\ref{Fig1}, that along with an increasing number of particles $N$ consecutive minima become deeper and maxima do not saturate at zero. This experimentally confirmed behavior (red dashed line in Fig.~\ref{Fig1}) is a direct manifestation of the collectivity of all particles in the system, which requires taking into account interactions between all possible pairs of opposite-spin fermions. As explained in \cite{2016HofmannPRA}, reduction of the problem to pure intra-shell pairing interactions (meaning that the system is treated as a collection of independent bounded pairs) leads to the significantly different behavior of the separation energy and consequently to evident contradiction with experimental results (see Fig.~\ref{Fig2}). 

Further analysis of the {\it even-odd} behavior can be continued even more deeply by considering not only the ground-state energy but also low-lying many-body excited states. In this way, it is possible to extract a few-body counterpart of the pairing gap well-known from the bulk BCS theory and compare their features \cite{2015DAmicoPRA}. It turns out that both quantities significantly differ not only in values but also in their functional dependence on interaction strength, even in the perturbative limit of vanishing interactions. These obvious discrepancies shall be attributed to a fundamental distinctness of mesoscopic few-fermion systems being well-described by perturbative means and many-body bulk systems whose pairing properties are rigorously non-perturbative (for more details see {\it e.g.} \cite{1971FetterBook,1997MatveevPRL,2016HofmannPRA}). 

It should be emphasized at this point that the energetic analysis, although very useful and well-incorporated by state-of-the-art experiments, has rather limited usefulness when more complicated questions and scenarios are posted. For example, giving decisive arguments for or against the formation of unconventional pairing in mass- or particle-imbalanced systems is far beyond the capacity of the method. Therefore other, more detailed studies based on inter-particle quantum correlations have to be considered. 

\begin{figure}
\includegraphics[width=\linewidth]{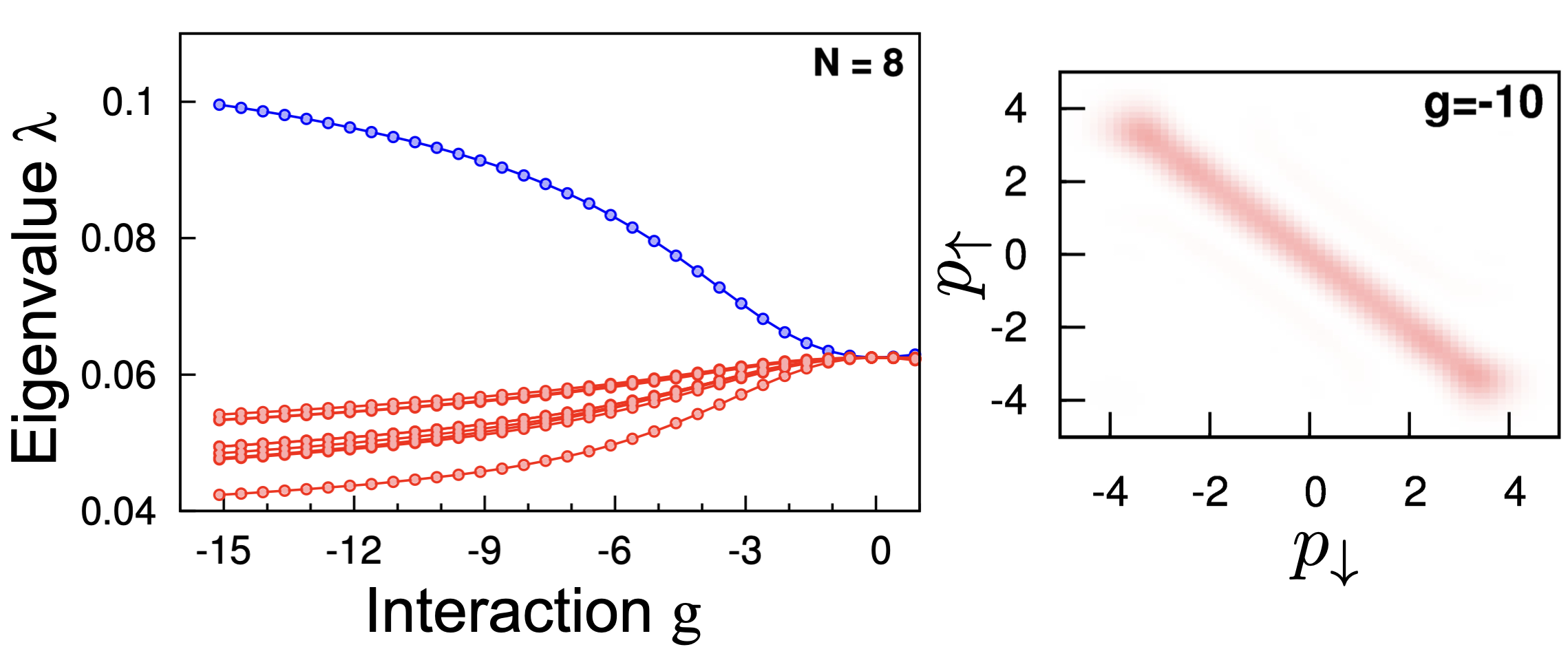}
\caption{{\bf (left)} Spectrum of the reduced two-particle density matrix \eqref{2rdm} as functions of strength of attractive interactions for the balanced system of $N_\uparrow=N_\downarrow=4$ equal mass fermions confined in a harmonic trap ($V_\sigma(x)=m\Omega^2x^2/2$). It is clear that along with increasing interactions one of the eigenvalues start to dominate in the system. This signals (partial) condensation of opposite-spin pairs to the dominant eigenorbital. {\bf (right)} Two-particle momentum distribution $|\widetilde\varphi_0(p_\uparrow,p_\downarrow)|^2$ encoded in the dominant eigenorbital obtained for very strong attractions. Note almost perfect anti-correlation of momenta in the most probable configurations accompanied by an almost uniform distribution of individual momenta. Interactions and momenta are expressed in natural units of harmonic oscillator, $\sqrt{\hbar^3\Omega/m}$ and $\sqrt{\hbar m\Omega}$, respectively. Figure based on \cite{2015SowinskiEPL} published by the IOP Publishing Ltd. under the terms of the Creative Commons Attribution 3.0 License.\label{Fig3}}
\end{figure}

{\bf Condensation of pairs.} -- Pairing between opposite-spin fermions forced by attractive interactions, independently of its particular structure, is a correlation phenomenon which in principle should be detectable by appropriately designed two-body measurement. Therefore, it must be somehow encoded in the most general quantity encoding all possible two-body correlations -- the reduced two-particle density matrix \cite{2017RammelPRA}. In the position domain it has the form:
\begin{multline} \label{2rdm}
\rho(x_\uparrow,x_\downarrow;x_\downarrow',x_\uparrow') = \\ \frac{1}{N_\uparrow N_\downarrow} \langle \mathtt{g}|\hat\psi^\dagger_\uparrow(x_\uparrow)\hat\psi^\dagger_\downarrow(x_\downarrow)\hat\psi_\downarrow(x'_\downarrow)\psi_\uparrow(x'_\uparrow)|\mathtt{g}\rangle.
\end{multline}
In generic situations, transition to the BCS phase is associated with Bose-Einstein condensation of Cooper pairs to single, well-defined two-particle state. It means that in the ideal scenario, when the system were composed only of identical pairs, the matrix \eqref{2rdm} would be expressed as a simple projector $\rho_{\mathrm{ideal}}(x_\uparrow,x_\downarrow;x_\downarrow',x_\uparrow') = \varphi_0(x_\downarrow,x_\uparrow)\varphi^*_0(x'_\downarrow,x'_\uparrow)$. Then, the two-particle orbital $\varphi_0(x_\uparrow,x_\downarrow)$ is interpreted as a two-particle wave function describing any two opposite-spin fermions in the BCS phase. Of course in any realistic situation the density matrix is much more complicated. However, if indeed the condensation of pairs to a selected orbital happens, then it must be reflected in the spectral decomposition of the reduced density matrix. Namely, if the reduced density matrix \eqref{2rdm} is diagonalized and expressed as a sum of projectors on its two-particle eigenorbitals $\varphi_i(x_\uparrow,x_\downarrow)$ as
\begin{equation} \label{decomposition}
\rho(x_\uparrow,x_\downarrow;x_\downarrow',x_\uparrow') = \sum_i \lambda_i \varphi_i(x_\uparrow,x_\downarrow)\varphi^*_i(x'_\uparrow,x'_\downarrow)
\end{equation}
then one of the eigenorbitals $\varphi_0(x_\uparrow,x_\downarrow)$ has a significantly larger eigenvalue $\lambda_0$. Since all the eigenvalues sum up to one, $\sum_i \lambda_i=1$, $\lambda_0$ can be interpreted as a fraction of Cooper-paired fermions in the system. In principle, this path of exploration can be adapted for any few-fermion system. In the case of an ideally balanced system ($N_\uparrow=N_\downarrow$) of equal-mass fermions confined in harmonic trap  
the situation is very clear \cite{2015SowinskiEPL}. As shown in Fig.~\ref{Fig3}, when attractions between opposite-spin fermions are increased one of the eigenvalues starts to dominate in the decomposition \eqref{decomposition}. Moreover, the corresponding eigenorbital in the momentum domain, $\widetilde\varphi_0(p_\uparrow,p_\downarrow)=\int dx_\uparrow dx_\downarrow\,\exp(-ip_1x_\uparrow-ip_2x_\downarrow)\varphi_0(x_\uparrow,x_\downarrow)$, displays correlations specific for Cooper pairs, {\it i.e.}, correlated fermions have almost exactly opposite momenta, $p_\uparrow+p_\downarrow=0$, while their individual momenta are rather not well-determined and they are distributed among all accessible values. 

In the case of imbalanced systems, the interpretation of corresponding results is no longer straightforward. Even if one considers only small deviations from perfect particle balance ($N_\uparrow-N_\downarrow=1$), instead of one well-distinguished dominant orbital one finds two different orbitals having similar weights whose roles in collective pairing are not well-clarified \cite{2015SowinskiFBS}. Since there is no experimental way to distinguish co-existing orbitals, providing any phenomenological interpretation to these results may be misleading.

\begin{figure}
\includegraphics[width=\linewidth]{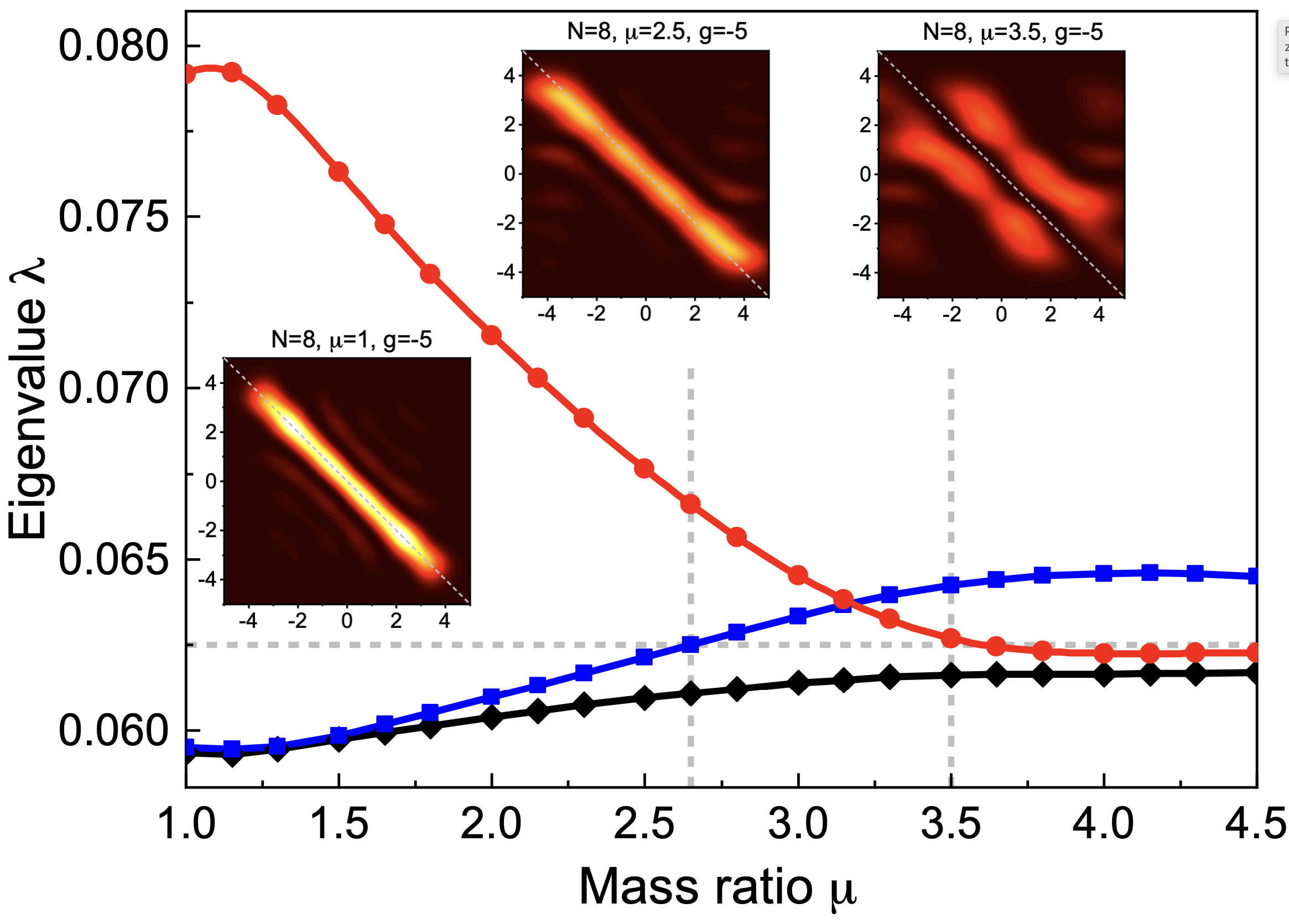}
\caption{Spectrum of the reduced two-particle density matrix \eqref{2rdm} for the balanced system of attractively interacting $N_\uparrow=N_\downarrow=4$ fermions as a function of their mass ratio $\mu=m_\downarrow/m_\uparrow$ and confined in a harmonic trap ($V_\sigma(x)=m_\sigma\Omega^2x^2/2$). Along with increasing mass ratio, the dominant eigenvalue decreases and around $\mu\approx 2.7$ the second eigenvalue becomes larger than its noninteracting value (horizontal dashed line). For larger mass ratios, two-particle orbitals exchange their domination and other pairing mechanism (with other momentum correlations) starts to determine system's properties. {\bf (insets)} Two-particle momentum distributions of opposite-spin fermions occupying dominant orbital for three different mass ratios $\mu=1$, $\mu=2.5$, and $\mu=3.5$. Note a significant enhancement of non-zero net momentum in the latter case signalling the non-trivial FFLO pairing phase. In all plots interaction strength and momenta are expressed in natural units of the problem, $\sqrt{\hbar^3\Omega/m_\uparrow}$ and $\sqrt{\hbar m_\uparrow\Omega}$, respectively. Figure adapted with permission from \cite{2020LydzbaPRA}. Copyright (2020) by the American Physical Society.\label{Fig4}}
\end{figure}

The situation is better if, instead of particle imbalance, mass imbalance is considered. Then we deal with the continuous variable $\mu=m_\downarrow/m_\uparrow$ and therefore we can precisely follow the system's properties when the parameter is varied and finally capture the moment when the description in terms of a single dominant orbital breaks down \cite{2020LydzbaPRA}. Fig.~\ref{Fig4} collects the most important results for this case obtained for a particular attraction strength ($g=-5$ in chosen dimensionless units). When the mass ratio $\mu$ is not far from unity the system resembles a mass-balanced system with a single orbital dominating the two-particle density matrix. It is characterised by zero net momentum pairing, $p_\uparrow+p_\downarrow=0$. Along with increasing mass imbalance, domination of the orbital is diminished while another orbital becomes more prominent. For sufficiently large mass imbalance the roles of these orbitals are reversed. Importantly, for the latter orbital, we note enhancement of non-zero net momentum in the two-particle distribution. This indicates the appearance of the FFLO phase rather than standard BCS pairing. 

{\bf Accessible two-body correlations.} --
One of the most important advantages of the approach based on the reduced density matrix presented above is that it gives direct and complete access to any possible two-particle correlations. From the experimental point of view, however, it is very challenging, if possible at all, to measure this highly non-trivial quantity. In fact, it requires measuring not only instantaneous correlations between the particles' positions or momenta but also all off-diagonal terms responsible for two-particle coherence. In consequence, experimental verification of whether the two-particle reduced density matrix contains dominant orbitals, and exposition of their internal structures are almost not possible in practice. On the other hand, it is clear (as we mentioned previously) that focusing only on global properties of the system, like ground-state energies, density profiles, {\it etc.}, cannot be sufficient to capture the essential features of pairing being a collective and purely quantum property encoded in the inter-particle correlations. Therefore, it is important to find some intermediate approach which on one hand is experimentally feasible, while on the other gives conclusive tools for capturing pairing phases.

One of the possible solutions is to consider not the whole two-particle density matrix \eqref{2rdm} but only those of its parts which are experimentally accessible. In this context, the best candidates are diagonal parts of the matrix $\rho(x_\uparrow,x_\downarrow;x_\downarrow',x_\uparrow')$ when expressed in the position or momentum domain. These parts correspond directly to two-particle density distributions $n^{(2)}(x_\uparrow,x_\downarrow)$ and $\widetilde{n}^{(2)}(p_\uparrow,p_\downarrow)$ describing probability of simultaneous finding of opposite-spin fermions at positions $x_\uparrow,x_\downarrow$ or with momenta $p_\uparrow,p_\downarrow$. It is clear that these quantities encode only some parts of all possible two-particle correlations. Therefore, the information gain from their analysis cannot exceed the information obtained from a whole two-particle matrix. However, these correlations are the most important from a physical point of view as being directly accessible in experiments. From this perspective, the lack of direct evidence for pairing correlations in these densities would probably mean that performing experimental verification will remain beyond present possibilities. 

Since opposite-spin pairing is mainly manifested in the momentum domain, let us focus on the distribution $\widetilde{n}^{(2)}(p_\uparrow,p_\downarrow)$. It is clear that even in a purely non-interacting regime this distribution displays some correlations between fermions. They are caused by accidental coincidences of events occurying independently in both components. Indeed, in this limit the distribution is a direct product of independent single-particle distributions, $\widetilde{n}^{(2)}(p_\uparrow,p_\downarrow)=\widetilde{n}(p_\uparrow)\widetilde{n}(p_\downarrow)$. Therefore, to emphasise non-trivial correlations caused solely by interactions it is better to consider a slightly modified distribution -- {\it the noise correlation} -- originally introduced in \cite{2004AltmanPRA} and later applied for one-dimensional \cite{2008MatheyPRL} and few-body systems \cite{2017BrandtPRA}. In the scenario studied and in the momentum domain it is defined as
\begin{equation}
{\cal G}(p_\uparrow,p_\downarrow) = \widetilde{n}^{(2)}(p_\uparrow,p_\downarrow)-\widetilde{n}(p_\uparrow)\widetilde{n}(p_\downarrow).
\end{equation}
This distribution directly encodes genuine two-body correlations, {\it i.e.}, all those correlations which cannot be recovered from corresponding single-particle densities. Quite recently it was argued that in the case of bulk systems confined in optical lattices or flat box potentials, this distribution nicely captures the appearance of the FFLO phase in spin-imbalanced fermionic mixtures \cite{2008LuscherPRA,2020RammelSciPost}. Therefore, this is one of the natural candidates for detecting unconventional pairing phases also in the few-body regime.

\begin{figure}
\includegraphics[width=\linewidth]{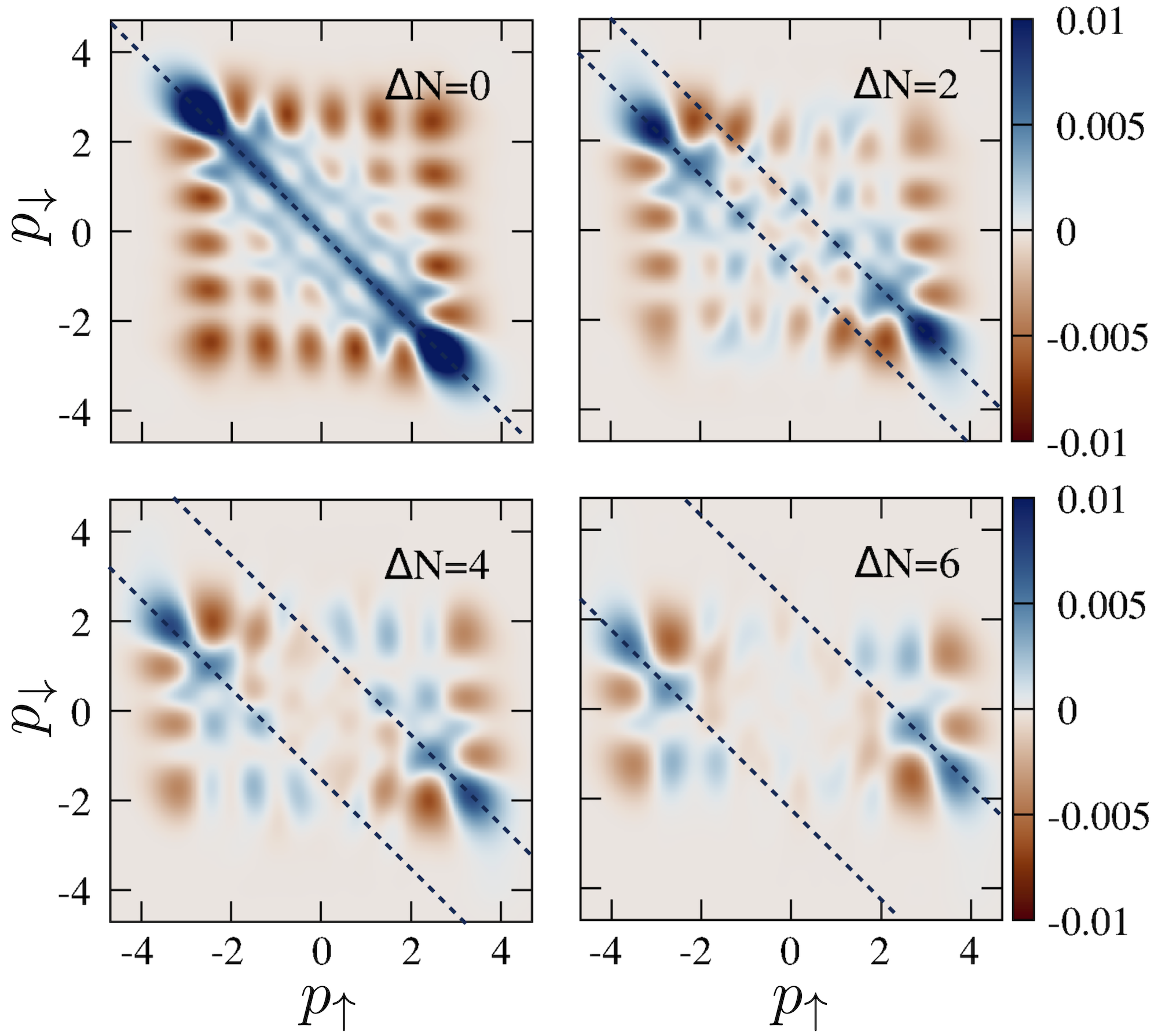}
\caption{Mutual correlations revealed in the noise correlation between fermions of opposite components in the momentum domain for the system of $N=10$ particles and different imbalances $\Delta N=N_\uparrow-N_\downarrow$. For the ideally balanced case ($\Delta N=0$), excellent anti-correlation of the momenta is visible signaling the emergence of the Cooper-like pairing. For imbalanced systems ($\Delta N\neq0$), opposite-spin fermions preferably form pairs having non-vanishing center-of-mass momentum $Q$. This can be viewed as a precursor of FFLO-like phase formation. Figure based on \cite{2020PecakPRR} published by the American Physical Society under the terms of the Creative Commons Attribution 4.0 International License. \label{Fig5}}
\end{figure}

\begin{figure}
\includegraphics[width=\linewidth]{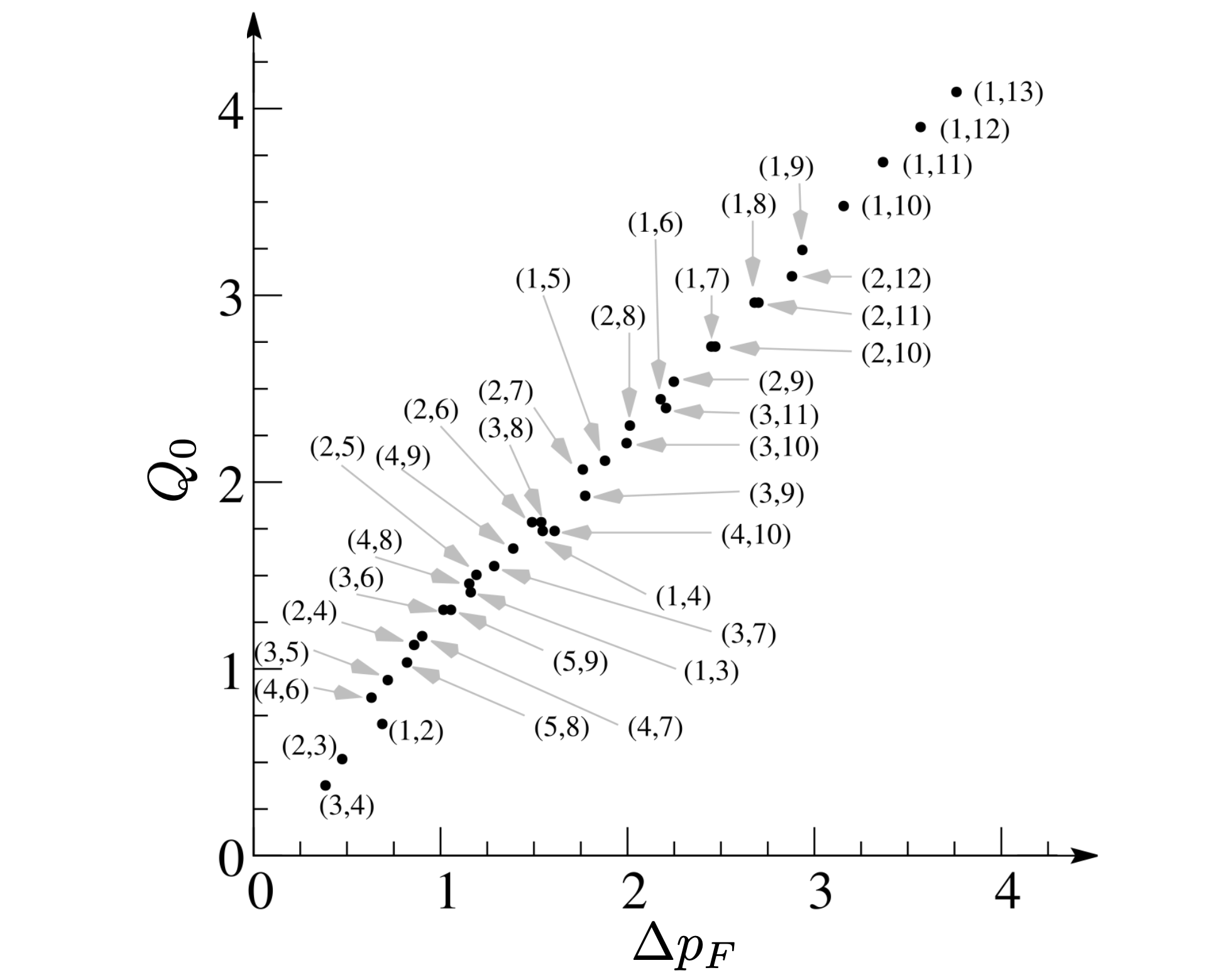}
\caption{The most probable center-of-mass momentum $Q_0$ for attractively interacting systems of different particle numbers $N_\uparrow$ and $N_\downarrow$ (indicated in parenthesis) as a function of the mismatch between Fermi momenta of both components $\Delta p_F$. Note an almost ideal linear dependence of these quantities. Figure reprinted from \cite{2020PecakPRR} published by the American Physical Society under the terms of the Creative Commons Attribution 4.0 International License. \label{Fig6}}
\end{figure}

It turns out that adaptation of the noise correlations distribution to capturing conventional and unconventional pairings works almost ideally in the case of systems with particle imbalance. Taking as a working example the case of harmonically trapped particles ($V_\sigma(x)=m\Omega^2 x^2/2$), different systems with up to $N=14$ particles and variety of imbalances were studied in \cite{2020PecakPRR}. It turned out (see Fig.~\ref{Fig5}) that the noise correlation ${\cal G}(p_\uparrow,p_\downarrow)$ displays clear evidence that in the case of an ideally balanced scenario ($\Delta N=N_\uparrow-N_\downarrow=0$) opposite-spin fermions are strongly anti-correlated in the momentum domain (note strong enhancement of the probability that opposite-spin fermions have exactly opposite momenta, $p_\uparrow=-p_\downarrow$). Moreover, when the particle imbalance ($\Delta N\neq0$) is introduced, the antidiagonal enhancement is split into two parts which are pushed out from the line $p_\uparrow+p_\downarrow=0$. Importantly, the most probable shift of the center-of-mass momentum $Q_0$ (dashed lines in Fig.~\ref{Fig5}) increases monotonically with the imbalance $\Delta N$ and it matches almost exactly the momentum difference $\Delta p_F$ which can be associated with a mismatch between Fermi momenta of both components confined in a harmonic trap (see Fig.~\ref{Fig6} and \cite{2020PecakPRR} for details). Obtaining this highly non-trivial prediction requires filtering out the most probable net momentum $Q_0$ from the noise correlation distribution. This can be done in many different ways, but all of them (if reasonable) give the same qualitative and quantitative result. The simplest way is to convolute the noise correlation function ${\cal G}(p_\uparrow,p_\downarrow)$ with a Gaussian filter of the center-of-mass momentum (see \cite{2020PecakPRR,2021DobrzynieckiArxiv} for details). 

Detection of an unconventional pairing between spin-imbalanced fermions based on the noise correlation as described above is very general and can be successfully utilized for different, even highly non-trivial confinements. As shown recently in \cite{2021DobrzynieckiArxiv}, by the appropriate introduction of the internal potential barrier experienced solely by one of the components one can convert the system between different FFLO phases characterized by a different center-of-mass momentum $Q_0$ without changing the particle imbalance. In this way, another path for experimental observation of unconventional pairings phases is provided. 
 
Unfortunately, in the case of mass-imbalanced few-fermion systems the situation is not as promising. For such systems, in contrast to the dominant orbital method, the noise correlation cannot be easily used to track unconventional pairing mechanisms. As shown in \cite{2019PecakPRA}, increasing the mass ratio $\mu$ does not force the noise correlation to signal the FFLO phase (as argued previously) but rather simply diminishes all opposite-spin correlations encoded in the diagonal part of the two-particle reduced density matrix. It means that all remaining correlations are rather forced by two-particle off-diagonal order -- although they are visible as small dominations of the standard or the FFLO eigenorbitals, they are almost undetectable in terms of noise correlation. This result simply reflects the fact that along with increasing mass ratio all eigenvalues become very close to their noninteracting values (see Fig.~\ref{Fig4}). This apparent contradiction shows, on the one hand, a limit of  the capabilities of the noise correlation and on the other the experimental inadequacy of the approach based on eigenorbitals of the reduced density matrix. 

{\bf Final remarks.} -- Spontaneous emergence of collectively enhanced pairing between opposite-spin fermions forced by attractive mutual interactions is one of the most spectacular, although quite common, quantum effects. For over 100 years it has continuously given rise to a variety of different applications in bulk solid-state systems. Recent experimental progress with ultracold atoms has opened a completely new possibility of studying this astonishing phenomenon in systems containing only mesoscopic number of particles. Currently, one can examine how the many-body collectiveness is mysteriously interlocked with the system when the number of particles is sequentially increased. In this perspective letter, we overviewed some theoretical concepts especially well tailored for one-dimensional systems, indicating an appropriate path for experimental confirmation of the existence of standard (BCS) as well as unconventional (FFLO) pairing phases.  

\acknowledgements
The author is extremely grateful to Jacek Dobrzyniecki for his comments and suggestions on the final draft of the manuscript. This work was supported by the (Polish) National Science Centre Grant No. 2016/22/E/ST2/00555.

\bibliographystyle{eplbib}
\bibliography{_biblio.bib}

\end{document}